\documentclass[amsmath,nofootinbib,aps,preprint]{revtex4}

\usepackage{bm}
\usepackage{amsmath}

\begin{document}

\title{Nonlinear $\sigma$-Model in (2+1) dimensions}
\author{Hao Huan}
 \email{hhuan@uchicago.edu} \affiliation{Department of Physics, University of Chicago, Chicago,
Illinois 60637}

\date{May 2008}
\begin{abstract}
    The nonlinear $\sigma$-model in (2+1) dimensions admits
    topological configurations called skyrmions. The topological
    charge of skyrmions turn out to be the fermionic number and the
    fermionic current is dictated by the skyrmion field
    configuration. The peculiar feature of this model is that a Hopf
    term introduced into the Lagrangian can lead to exotic spin and
    fractional statistics.
\end{abstract}
\maketitle

\section{Topological Foundation}\label{sec:topology}
    Consider a scalar field theory with a continuous symmetry group
    $G$. If the potential has a minimum which spontaneously break
    the symmetry to its subgroup $H$, there will be Goldstone modes
    corresponding to the broken generators, with the effective
    action\cite{Weinberg:1996kr}
    \begin{equation}
        S[\pi]=\int d^{d+1} x [\frac{1}{2}g_{ab}(\pi)\partial_i \pi_a
        \partial_i \pi_b+\cdots]\label{eq:goldstone}
    \end{equation}
    where $d$ is the dimension of the Euclidean space, $g_{ab}$ is
    a positive-definite metric, and there may be terms of higher
    order derivatives of $\pi$. For the gradient energy to be finite,
    $\partial_i\pi_a(\mathbf{x})$ should vanish at infinity faster
    than $|\mathbf{x}|^{-d/2}$, that is, $\pi_a(\mathbf{x})$ should
    approach a constant at infinity.

    The field $\pi_a(\mathbf{x})$ acts as a map from the
    $d$-dimensional Euclidean space toward the space of the
    Goldstone modes, i.e. the coset manifold $G/H$. Further,
    following the requirement of finite energy discussed above, the
    sphere at $|\mathbf{x}|\rightarrow \infty$ is mapped as one
    point. In other words, the $d$-dimensional space is compactified
    as a sphere $S_d$ and mapped to $G/H$ by the field
    configuration.

    The set of all the configurations is thus
    divided into several classes which are not connected to each
    other by a continuous path, according to the degree of the map.
    These are field configurations with different topology,
    corresponding to different elements in the $d$-th homotopy group
    of $G/H$. Moreover since they are not connected, configurations
    in one class cannot be slowly deformed to those in another
    class, e.g. by low temperature fluctuations, adiabatic time
    evolutions, etc. Hence the degree of the map is in most cases a
    conserved number.

    In this way we have found several static field configurations
    other than the trivial vacuum. The configurations with the
    lowest gradient energy in each class, because they are by
    definition at the local minimum of the potential energy
    everywhere, will also satisfy the action principle and so make
    themselves classical solutions to the theory. Because the
    degree of the map is topologically protected, quantum
    fluctuations can hardly destroy these configurations. Hence we
    expect them to appear in the quantized version of the scalar
    field theory too. The nonlinear $\sigma$ model as a specific
    example is discussed in the next section.

\section{Nonlinear $\sigma$-model}\label{sec:sigma}
    The nonlinear $\sigma$-model was first constructed to study the
    axial current in $\beta$-decays in the physical spacetime with (3+1) dimensions\cite{GellMann:1960np}. The Lagrangian is
    \begin{equation}
        \mathcal{L}=-\bar{N}[\gamma^{\mu}\partial_{\mu}+m_0-g_0(\sigma+i\boldsymbol{\tau}\cdot\boldsymbol{\pi}\gamma_5)]N-\mathcal{L}(\sigma,\boldsymbol{\pi})\label{eq:gellmann}
    \end{equation}
    where $N$ is the nucleon field,
    $\mathcal{L}(\sigma,\boldsymbol{\pi})$ is a function of the meson
    fields $\sigma$, $\boldsymbol{\pi}$ with the ordinary gradient term
    and a potential which acquires its local minimum at
    \begin{equation}
        \boldsymbol{\pi}^2+\sigma^2=C^2\label{eq:nonlinear}
    \end{equation}
    where $C$ is a constant. Now $\boldsymbol{\pi}$ and $\sigma$ are
    related by a nonlinear equation and the original symmetry group
    $SO(4)$ is broken to $SO(3)$ when a vacuum expectation value is
    selected. The topological classes of the meson field
    configuration is thus shown by the homotopy group $\pi_3(S_3)$,
    since $S_3=SO(4)/SO(3)$. The group is known to be the infinite
    cyclic group $\boldsymbol{Z}$, which tells that there are
    infinite topological classes, characterized by the conserved
    topological charge $Q=0,\pm 1, \pm 2, \cdots$ The configuration
    with $Q=1$ is called a skyrmion\cite{Skyrme:1962vh}.

    For reasonable interest here we consider the nonlinear
    $\sigma$-model in (2+1) dimensions instead. The homotopy group
    concerned now is $\pi_2(G/H)$ and for the skyrmion to appear, we
    want it to be the infinite cyclic group as well. A natural
    candidate is to choose $G=SO(3)$, $H=SO(2)$, and
    $G/H=SO(3)/SO(2)=S_2$. Therefore we need to have three
    components for the scalar field and couple them to the fermion
    field. Resembling the $\sigma$-model in (3+1) dimensions, the
    model is constructed using the Yukawa coupling
    \begin{equation}
        \mathcal{L}=\bar{\psi}(i\gamma^{\mu}\partial{\mu}-m\boldsymbol{n}\cdot\boldsymbol{\sigma})\psi-\mathcal{L}(\boldsymbol{n})\label{eq:2d}
    \end{equation}
    where $\boldsymbol{\sigma}$ are the Pauli matrices, and the
    potential of the $\boldsymbol{n}$ field achieves its local minimum
    when $\boldsymbol{n}$ is a unit vector. Without the fermion
    part, the Lagrangian can be regarded as describing a ferromagnet
    in (2+1) dimensions with $\boldsymbol{n}$ as the spin vector.

    There are some notes about the model construction. Because the
    spacetime is (2+1) dimensional, the Clifford algebra is closed,
    which is just the algebra of the three Pauli matrices. Since we
    also have three scalar field components to couple here, there is
    no $\gamma_5$ term possible in the Lagrangian. Also we have
    assumed the fermion mass $\mu$ to be zero, or just $\mu<<m$.

    $\pi_2(S_2)=\boldsymbol{Z}$ tells the existence of the topological
    configurations, i.e. the nontrivial metastable states. The next
    section will give the explicit expression for these classical
    solutions and their energy spectrum as well.

\section{Skyrmion Solution}\label{sec:skyrmion}
    For any field configuration in the (2+1) dimensional nonlinear
    $\sigma$-model, the degree of the map, or the topological
    charge,
    is expressed as\cite{Polyakov:1975yp}
    \begin{equation}
        Q=\frac{1}{8\pi}\int d^2x
        \epsilon^{\mu\nu}\epsilon^{abc}n_{a}\partial_{\mu}n_{b}\partial_{\nu}n_c\label{eq:degree}
    \end{equation}
    where $\mu,\nu=1,2$ and $a,b,c=1,2,3$. It has a simple form in
    spherical coordinates $n_1=\sin \theta\cos\phi$, $n_2=\sin
    \theta\sin\phi$, $n_3=\cos \theta$
    \begin{equation}
        Q=\frac{1}{4\pi}\int
        \sin\theta(x)d\theta(x)d\phi(x)\label{eq:spherical}
    \end{equation}
    from which it is clear that $Q$ is an integer and is just the
    number of times the sphere $S_2$ is wrapped around the unit
    sphere of $\boldsymbol{n}$.

    Now we have the inequality
    \begin{equation}
        (\partial_{\mu}n^a +
        \epsilon_{\mu\nu}\epsilon^{abc}n_b\partial^{\nu}n_c)^2\geq
        0\label{eq:oineq}
    \end{equation}
    which is expanded to be
    \begin{equation}
        \partial_{\mu}n_a\partial^{\mu}n^a
        +2\epsilon^{\mu\nu}\epsilon^{abc}\partial_{\mu}n_a
        n_b\partial_{\nu}n_c+\epsilon^{abc}\epsilon_a^{\text{  }b'c'}\epsilon^{\mu\nu}\epsilon_{\mu}^{\text{  }\nu'}n_b\partial_{\nu}n_c
        n_{b'}\partial_{\nu'}n_{c'}\geq 0\label{eq:expand}
    \end{equation}
    Using the formulae
    \begin{eqnarray}
        \epsilon^{abc}\epsilon_a^{\text{
        }b'c'}&=&\delta^{bb'}\delta^{cc'}-\delta^{bc'}\delta^{b'c}\\\label{eq:eq1}
        \epsilon^{\mu\nu}\epsilon_{\mu}^{\text{
        }\nu'}&=&g^{\nu\nu'}\label{eq:eq2}
    \end{eqnarray}
    it is found
    \begin{equation}
        \partial_{\mu}n_a\partial^{\mu}n^a\geq
        \epsilon^{\mu\nu}\epsilon^{abc}n_{a}\partial_{\mu}n_{b}\partial_{\nu}n_c\label{eq:rearrange}
    \end{equation}
    Since the gradient energy
    \begin{equation}
        E=\int d^2x
        \partial_{\mu}n_a\partial^{\mu}n^a\label{eq:energy}
    \end{equation}
    the definition in Eq.~\ref{eq:degree} gives
    \begin{equation}
        E\geq 8\pi Q\label{eq:bound}
    \end{equation}
    which is the lower bound of the energy in each homotopy class,
    i.e. the energy of the metastable states.

    The above discussion also tells the static solutions satisfy
    \begin{equation}
        \partial_{\mu}n^a=-\epsilon_{\mu\nu}\epsilon^{abc}n_b\partial^{\nu}n_c\label{eq:solution}
    \end{equation}
    Introducing a complex variable $w$, where
    \begin{eqnarray}
        w&=&w_1+iw_2\\\label{eq:component}
        w_1&=&\cot\frac{\theta}{2}\cos\phi\\\label{eq:w1}
        w_2&=&\cot\frac{\theta}{2}\cos\phi\label{eq:w2}
    \end{eqnarray}
    the condition gives
    \begin{eqnarray}
        \frac{\partial w_1}{\partial x_1}&=&\frac{\partial
        w_2}{\partial x_2}\\\label{eq:firstcr}
        \frac{\partial w_2}{\partial x_1}&=&-\frac{\partial
        w_1}{\partial x_2}\label{eq:secondcr}
    \end{eqnarray}
    which are recognized to be the Cauchy-Riemann conditions for a
    holomorphic function in the complex plane $z=x_1+ix_2$.

    To find the detailed form of the function $w(z)$, the boundary
    condition is needed. As the discussion in
    Sec.~\ref{sec:topology} tells, the field should approach a
    constant at $|z|\rightarrow\infty$. Because of the global
    symmetry, we can assume the asymptotic value to be
    $\boldsymbol{n}\rightarrow(0,0,1)$, that is, $\theta$ is zero at
    infinity. The field $\boldsymbol{n}$ is a continuous function of
    the coordinates, so the only singularities possible are poles.
    Now we can write down the solution
    \begin{equation}
        w(z)=\prod_i(\frac{z-z_i}{\lambda})^{m_i}\prod_j(\frac{\lambda}{z-z_j})^{n_j}\label{eq:complex}
    \end{equation}
    where the boundary condition requires
    \begin{equation}
        \sum_i m_i>\sum_j n_j\label{eq:condition}
    \end{equation}
    According to Eq.~\ref{eq:spherical}, the topological charge $Q$
    is the number of solutions of Eq.~\ref{eq:complex} that
    express $z$ in terms of $w$, i.e.
    \begin{equation}
        Q=\sum_i m_i\label{eq:relation}
    \end{equation}

    As an example let's examine the skyrmion case, that is, Q=1. The
    complex solution reduces to
    \begin{equation}
        w(z)=\frac{z-z_0}{\lambda}\label{eq:skyrmioncomplex}
    \end{equation}
    We can simply move the origin in the complex plane to $z_0$.
    Then using
    \begin{equation}
        w=\cot\frac{\theta}{2}e^{i\phi}\label{eq:w}
    \end{equation}
    we have
    \begin{eqnarray}
        \frac{n_1}{1-n_3}&=&\frac{x_1}{\lambda}\\\label{eq:firstskyr}
        \frac{n_2}{1-n_3}&=&\frac{x_2}{\lambda}\label{eq:secondskyr}
    \end{eqnarray}
    The explicit solutions are
    \begin{eqnarray}
        n_1&=&\frac{x_1}{r}\cdot\frac{2\frac{\lambda}{r}}{1+\frac{\lambda^2}{r^2}}\\\label{eq:n1}
        n_2&=&\frac{x_2}{r}\cdot\frac{2\frac{\lambda}{r}}{1+\frac{\lambda^2}{r^2}}\\\label{eq:n2}
        n_3&=&\frac{1-\frac{\lambda^2}{r^2}}{1+\frac{\lambda^2}{r^2}}\label{eq:n3}
    \end{eqnarray}
    where $r=\sqrt{x_1^2+x_2^2}$. Finally we find that it is just a
    stereographic projection, with $\lambda$ a scale factor.

    The existence of the scale factor is interesting. It means there
    are a family of metastable solutions for a given homotopy class.
    However we can expect the scale factor to be restricted by the
    higher order derivative terms in the effective Lagrangian for
    the $\boldsymbol{n}$ field, that is, a small perturbation in the
    skyrmion energy which depends on $\lambda$.

    Up till now we have explicitly shown that the static solutions
    with nontrivial topological charges indeed exist in the
    (2+1)-dimensional nonlinear $\sigma$-model. To understand the
    physical significance of the topological charge, besides
    affecting the configuration energy, the coupling to the fermion
    fields should be taken into consideration. In fact, as a result
    of the index theorem, the topological charge equals the index of
    the Hamiltonian, that is, the fermionic charge of the $\psi$
    field. To verify this relation we will calculate the vacuum
    expectation value of the fermionic current in the presence of
    the topological configurations in the next section.

\section{Fermionic Current}\label{sec:current}
    We follow the path integral approach for the expectation value
    $\langle j^{\mu}\rangle$. That is, introducing an auxiliary
    field $A_{\mu}$, we rewrite the Lagrangian as
    \begin{equation}
        \mathcal{L}=\bar{\psi}(i\gamma^{\mu}\partial_{\mu}+\gamma^{\mu}A_{\mu}-m\boldsymbol{n}\cdot\boldsymbol{\sigma})\psi\label{eq:auxiliary}
    \end{equation}
    where the fluctuations in the $\boldsymbol{n}$ field is ignored.
    Therefore
    \begin{eqnarray}
        \langle j^{\mu}\rangle&=&\frac{\int
        \mathcal{D}\bar{\psi}\mathcal{D}\psi
        \bar{\psi}\gamma^{\mu}\psi\exp(i\int d^3x
        \mathcal{L})}{\int
        \mathcal{D}\bar{\psi}\mathcal{D}\psi
        \exp(i\int d^3x
        \mathcal{L})}|_{A_{\mu}=0}\\\label{eq:origin}
        &=&-i\frac{\frac{\delta}{\delta A_{\mu}}\int
        \mathcal{D}\bar{\psi}\mathcal{D}\psi
        \exp(i\int d^3x
        \mathcal{L})}{\int
        \mathcal{D}\bar{\psi}\mathcal{D}\psi
        \exp(i\int d^3x
        \mathcal{L})}|_{A_{\mu}=0}\label{eq:delta}
    \end{eqnarray}
    that is,
    \begin{eqnarray}
        \langle j^{\mu}\rangle&=&-i\frac{\frac{\delta}{\delta
        A_{\mu}}e^{iS_{\text{eff}}[A_{\mu},\boldsymbol{n}]}}{e^{iS_{\text{eff}}[A_{\mu},\boldsymbol{n}]}}|_{A_{\mu}=0}\\\label{eq:effaction}
        &=&\frac{\delta
        S_{\text{eff}}[A_{\mu},\boldsymbol{n}]}{\delta
        A_{\mu}}|_{A_{\mu}=0}\label{eq:deltaeffaction}
    \end{eqnarray}
    where
    \begin{equation}
        e^{iS_{\text{eff}}[A_{\mu},\boldsymbol{n}]}=\int
        \mathcal{D}\bar{\psi}\mathcal{D}\psi
        \exp(i\int d^3x
        \mathcal{L})\label{eq:effactiondef}
    \end{equation}
    From the Grassmannian integral formula, we know
    \begin{eqnarray}
        e^{iS_{\text{eff}}[A_{\mu},\boldsymbol{n}]}&=&\int
        \mathcal{D}\bar{\psi}\mathcal{D}\psi
        \exp(i\int d^3x
        \bar{\psi}(i\gamma^{\mu}\partial_{\mu}+\gamma^{\mu}A_{\mu}-m\boldsymbol{n}\cdot\boldsymbol{\sigma})\psi)\\\label{eq:plugin}
        &=&\det
        (-i\gamma^{\mu}\partial_{\mu}-\gamma^{\mu}A_{\mu}+m\boldsymbol{n}\cdot\boldsymbol{\sigma})\label{eq:det}
    \end{eqnarray}
    so
    \begin{eqnarray}
        S_{\text{eff}}[A_{\mu},\boldsymbol{n}]&=&-i\log\det
        (-i\gamma^{\mu}\partial_{\mu}-\gamma^{\mu}A_{\mu}+m\boldsymbol{n}\cdot\boldsymbol{\sigma})\\\label{eq:log}
        &=&-i\text{tr}\log(-i\gamma^{\mu}\partial_{\mu}-\gamma^{\mu}A_{\mu}+m\boldsymbol{n}\cdot\boldsymbol{\sigma})\label{eq:trace}
    \end{eqnarray}
    Denoting
    $D=-i\gamma^{\mu}\partial_{\mu}-\gamma^{\mu}A_{\mu}+m\boldsymbol{n}\cdot\boldsymbol{\sigma}$,
    Eq.~\ref{eq:deltaeffaction} tells
    \begin{eqnarray}
        \langle j^{\mu}\rangle&=&-i\frac{\delta\text{tr}\log
        D}{\delta A{_\mu}}|_{A_{\mu}=0}\\\label{eq:currentlog}
        &=&-i\text{tr}(\frac{\delta D}{\delta
        A_{\mu}}D^{-1})|_{A_{\mu}=0}\\\label{eq:currenttr}
        &=&-i\text{tr}(-\gamma^{\mu}D_0^{-1})\label{eq:currentgamma}
    \end{eqnarray}
    where
    \begin{eqnarray}
        D_0&=&D|_{A_{\mu}=0}\\\label{eq:d0}
        &=&-i\gamma^{\mu}\partial_{\mu}+m\boldsymbol{n}\cdot\boldsymbol{\sigma}\label{eq:d0detail}
    \end{eqnarray}
    Hence
    \begin{eqnarray}
        \langle
        j^{\mu}\rangle&=&i\text{tr}(\gamma^{\mu}D_0^{-1})\\\label{eq:currentsign}
        &=&i\text{tr}(\gamma^{\mu}D_0^{-1}D_0^{\dag
        -1}D_0^{\dag})\\\label{eq:currentdag}
        &=&i\text{tr}(\gamma^{\mu}(D_0^{\dag}D_0)^{-1}D_0^{\dag})\label{eq:currentinverse}
    \end{eqnarray}
    where
    \begin{eqnarray}
        D_0^{\dag}&=&i\gamma^{\mu}\partial_{\mu}+m\boldsymbol{n}\cdot\boldsymbol{\sigma}\\\label{eq:d0dag}
        D_0^{\dag}D_0&=&\partial^2+m^2+im\gamma^{\nu}\partial_{\nu}\boldsymbol{n}\cdot\boldsymbol{\sigma}\label{eq:d0d0dag}
    \end{eqnarray}
    that is
    \begin{equation}
        \langle
        j^{\mu}\rangle=i\text{tr}(\gamma^{\mu}\frac{1}{\partial^2+m^2+im\gamma^{\nu}\partial_{\nu}\boldsymbol{n}\cdot\boldsymbol{\sigma}}(i\gamma^{\rho}\partial_{\rho}+m\boldsymbol{n}\cdot\boldsymbol{\sigma}))\label{eq:currentfinal}
    \end{equation}
    To avoid confusions we use Greek letters for the spacetime
    indices and Latin letters for the internal space indices in the
    following calculation. The gradient expansion is used to
    evaluate the trace, that is
    \begin{equation}
        \frac{1}{\partial^2+m^2+im\gamma^{\nu}\partial_{\nu}\boldsymbol{n}\cdot\boldsymbol{\sigma}}=\frac{1}{\partial^2+m^2}(1+\sum_{k=1}^{\infty}(-im\gamma^{\nu}\partial_{\nu}n_{a}\sigma^{a}\frac{1}{\partial^2+m^2})^k)\label{eq:gradientexp}
    \end{equation}
    Since the scalar fields only act as a background, we can let
    $m\rightarrow\infty$, and determine the possible nonvanishing
    terms using dimensional analysis. The Lagrangian
    Eq.~\ref{eq:auxiliary} tells that the current $j^{\mu}$ is of
    mass dimension two, hence only the first three two terms in the
    gradient expansion, with 0, 1, and 2 derivatives respectively,
    may have a nonvanishing trace. We examine them one by one
    \begin{eqnarray}
        \langle
        j^{\mu}\rangle^{(0)}&=&i\text{tr}(\gamma^{\mu}\frac{1}{\partial^2+m^2}(i\gamma^{\rho}\partial_{\rho}+m\boldsymbol{n}\cdot\boldsymbol{\sigma}))\\\label{eq:zerothfirst}
        &=&-\text{tr}(\gamma^{\mu}\frac{1}{\partial^2+m^2}\gamma^{\rho}\partial_{\rho})+i\text{tr}(\gamma^{\mu}\frac{1}{\partial^2+m^2}m\boldsymbol{n}\cdot\boldsymbol{\sigma})\label{eq:zerothsecond}
    \end{eqnarray}
    where the first term vanishes because it is an odd function for
    the four-momentum of the eigenmodes, or in other words, an odd
    function of $\partial_{\mu}$. The second term is also zero since
    $\text{tr}\boldsymbol{\sigma}=0$. Now for the first order
    expansion term
    \begin{eqnarray}
        \langle
        j^{\mu}\rangle^{(1)}&=&\text{tr}(\gamma^{\mu}\frac{1}{\partial^2+m^2}m\gamma^{\nu}\partial_{\nu}n_{a}\sigma^{a}\frac{1}{\partial^2+m^2}(i\gamma^{\rho}\partial_{\rho}+m\boldsymbol{n}\cdot\boldsymbol{\sigma}))\\\label{eq:firstfirst}
        &=&i\text{tr}(\gamma^{\mu}\frac{1}{\partial^2+m^2}m\gamma^{\nu}\partial_{\nu}n_{a}\sigma^{a}\frac{1}{\partial^2+m^2}\gamma^{\rho}\partial_{\rho})\\&+&\text{tr}(\gamma^{\mu}\frac{1}{\partial^2+m^2}m\gamma^{\nu}\partial_{\nu}n_{a}\sigma^{a}\frac{1}{\partial^2+m^2}m\boldsymbol{n}\cdot\boldsymbol{\sigma})\label{eq:firstsecond}
    \end{eqnarray}
    where the first term vanishes for the same reason as in the
    zeroth expansion, and the second term can be rewritten as
    \begin{equation}
        \langle
        j^{\mu}\rangle^{(1)}=4m^2n_a\partial^{\mu}n^a\text{tr}(\frac{1}{\partial^2+m^2}\frac{1}{\partial^2+m^2})\label{eq:secondrearr}
    \end{equation}
    where we have used the following identities
    \begin{eqnarray}
        \text{tr}(\gamma^{\mu}\gamma^{\nu})&=&2g^{\mu\nu}\\\label{eq:gammaid}
        \text{tr}(\sigma^a\sigma^b)&=&2\delta^{ab}\label{sigmaid}
    \end{eqnarray}
    Hence $\langle j^{\mu}\rangle^{(1)}$ also vanishes because $\boldsymbol{n}$
    is a unit vector. The only possibility left is the
    second order expansion
    \begin{eqnarray}
        \langle
        j^{\mu}\rangle^{(2)}&=&-i\text{tr}(\gamma^{\mu}\frac{1}{\partial^2+m^2}m\gamma^{\nu}\partial_{\nu}n_{a}\sigma^a\frac{1}{\partial^2+m^2}m\gamma^{\rho}\partial_{\rho}n_{b}\sigma^b\frac{1}{\partial^2+m^2}(i\gamma^{\lambda}\partial_{\lambda}+m\boldsymbol{n}\cdot\boldsymbol{\sigma}))\\\label{eq:secondfirst}
        &=&\text{tr}(\gamma^{\mu}\frac{1}{\partial^2+m^2}m\gamma^{\nu}\partial_{\nu}n_{a}\sigma^a\frac{1}{\partial^2+m^2}m\gamma^{\rho}\partial_{\rho}n_{b}\sigma^b\frac{1}{\partial^2+m^2}\gamma^{\lambda}\partial_{\lambda})\\
        &-&i\text{tr}(\gamma^{\mu}\frac{1}{\partial^2+m^2}m\gamma^{\nu}\partial_{\nu}n_{a}\sigma^a\frac{1}{\partial^2+m^2}m\gamma^{\rho}\partial_{\rho}n_{b}\sigma^b\frac{1}{\partial^2+m^2}m\boldsymbol{n}\cdot\boldsymbol{\sigma})\label{eq:secondsecond}
    \end{eqnarray}
    Previous experiences tell us only the latter part needs to be
    considered. Using the identities
    \begin{eqnarray}
        \text{tr}(\gamma^{\mu}\gamma^{\nu}\gamma^{\rho})&=&-2i\epsilon{\mu\nu\rho}\\\label{eq:gammaide}
        \text{tr}(\sigma^a\sigma^b\sigma^c)&=&2i\epsilon^{abc}\label{eq:sigmaide}
    \end{eqnarray}
    we find
    \begin{eqnarray}
        \lim_{m\rightarrow\infty}\langle
        j^{\mu}\rangle&=&\langle
        j^{\mu}\rangle^{(2)}\\\label{eq:nonvanish}
        &=&-4im^3\epsilon^{\mu\nu\rho}\epsilon^{abc}n_a\partial_{\nu}n_b\partial_{\rho}n_c\text{tr}(\frac{1}{\partial^2+m^2}\frac{1}{\partial^2+m^2}\frac{1}{\partial^2+m^2})\label{eq:onlytrace}
    \end{eqnarray}
    where the trace of the differential operators can be evaluated
    as an integral in the three-dimensional Euclidean momentum space
    \begin{eqnarray}
        \text{tr}(\frac{1}{\partial^2+m^2}\frac{1}{\partial^2+m^2}\frac{1}{\partial^2+m^2})&=&i\int\frac{d^3\mathbf{p}}{(2\pi)^3}\frac{1}{(\mathbf{p}^2+m^2)^3}\\\label{eq:traceint}
        &=&\frac{i}{2\pi^2}\int_0^{\infty}\frac{p^2}{(p^2+m^2)^3}dp\label{eq:traceoned}
    \end{eqnarray}
    where $p=|\mathbf{p}|$. Since
    \begin{eqnarray}
        \int_0^{\infty}\frac{p^2}{(p^2+m^2)^3}dp
        &=&m^{-3}\int_0^{\infty}dx\frac{x^2}{(x^2+1)^3}\\\label{eq:intx}
        &=&m^{-3}\int_0^{\pi/2}d\theta
        \sin^2\theta\cos^2\theta\\\label{eq:inttheta}
        &=&\frac{\pi}{16}m^{-3}\label{eq:intresult}
    \end{eqnarray}
    we obtain the result for the expectation value of the fermionic
    current
    \begin{equation}
        \langle
        j^{\mu}\rangle=\frac{1}{8\pi}\epsilon^{\mu\nu\rho}\epsilon^{abc}n_a\partial_{\nu}n_b\partial_{\rho}n_c\label{eq:fermioncurrent}
    \end{equation}
    The charge associated with the current is just the fermionic
    number
    \begin{equation}
        F=\int d^2x \langle j^0\rangle\label{eq:fermionnumber}
    \end{equation}
    which is just the topological charge of the scalar field
    configuration defined in Eq.~\ref{eq:degree}. Therefore, we have
    confirmed the topological charge as the fermionic number of the
    coupled fermion field, integrated over all the fermion
    configurations. Physically speaking, in the presence of the
    topological configurations, fermions tend to dwell inside these
    scalar fields and carry a topologically protected fermionic
    number. The (2+1)-dimensional nonlinear $\sigma$-model provides
    a scenario for fermion number conservation.

    The above conclusions also apply to nonlinear $\sigma$-models in
    other dimensions. However, the (2+1)-dimension model has a
    unique feature which distinguishes itself, that the skyrmions
    may exhibit fractional statistics. This peculiar feature will be
    further explored in the next which is also the last section.

\section{Fractional Statistics}
    Even without coupling to the fermionic field, the topological
    configurations of the scalar may have their own statistical
    properties, maybe different from the scalar itself. In the
    (3+1)-dimensional nonlinear $\sigma$-model the skymion can be
    either a boson or a fermion, by introducing the
    Wess-Zumino-Witten term\cite{Witten:1983tx}. Here in our (2+1)-dimensional a similar
    term also exists, but leads to peculiar statistics of the
    skymion.

    To introduce the additional Hopf term\cite{Wilczek:1983cy} we notice that the
    topological current in Eq.~\ref{eq:fermioncurrent} naturally
    satisfies the current conservation equation
    \begin{equation}
        \partial_{\mu}j^{\mu}=0\label{eq:conserved}
    \end{equation}
    where we have dropped the bracket denoting the vacuum
    expectation value, since the current only depends on
    $\boldsymbol{n}$. Because the spacetime is three-dimensional, a
    gauge potential $A_{\mu}$ can be manufactured by the curl
    equation
    \begin{equation}
        j^{\mu}=\epsilon^{\mu\nu\lambda}\partial_{\nu}A_{\lambda}\label{eq:curl}
    \end{equation}
    where $j^{\mu}$ is invariant under the gauge transformation
    $A_{\mu}\rightarrow A_{\mu}-\partial_{\mu}\Lambda$.
    Eq.~\ref{eq:fermioncurrent} tells that $A_{\mu}$ is a functional
    of the $\boldsymbol{n}$ field with nonlocal dependence. The
    gauge field makes it possible to add to the action another term
    \begin{equation}
        S=S_0+\Theta H\label{eq:hopf}
    \end{equation}
    where $\Theta$ is an arbitrary parameter, and
    \begin{equation}
        H=-\frac{1}{2\pi}\int d^3x A_{\mu}j^{\mu}\label{eq:hopfdef}
    \end{equation}
    which is mathematically the homotopic invariant of the map
    $S_3\rightarrow S_2$. Just as the existence of the skyrmion is
    assured by $\pi_2(S_2)=\boldsymbol{Z}$, the integrity of
    introducing the topological Hopf term $H$ is due to the fact
    that $\pi_3(S_2)=\boldsymbol{Z}$.

    To discuss the statistics of the skyrmion affected by the Hopf
    term, we first realize that the statistics refers to the phase
    of interchanging two skyrmions, which is related to the phase of
    rotating one skyrmion by $2\pi$, that is, the spin. Therefore we
    consider the adiabatic process of rotating a topological
    configuration $\boldsymbol{n}(\mathbf{x})$ by the angle $2\pi$
    in the two-dimensional space, over a long period of time $T$.
    The spin or intrinsic angular momentum $J$ can be found by
    looking for the phase factor
    \begin{equation}
        e^{i2\pi J}=e^{iS}\label{eq:phase}
    \end{equation}
    where $S$ is the action corresponding to the adiabatic rotation
    process, consisting of the gradient part $S_0$ and the Hopf
    term. The gradient energy goes as $1/T^2$ and thus the action
    $S_0$ is of the order $1/T\rightarrow 0$ as $T\rightarrow
    \infty$. In other words, the skyrmion is a spin 0 scalar
    configuration without the Hopf term, as we expect.

    The effect of the Hopf term, it is convenient to recall that the
    three components of the $\boldsymbol{n}$ field satisfy the
    $SU(2)$ algebra and a rotation in space corresponds to an
    isospin rotation. In other words, we have a time-varying
    configuration $\boldsymbol{n}(\mathbf{x},t)$ which satisfies
    \begin{equation}
        \boldsymbol{n}(\mathbf{x},0)=\boldsymbol{n}(\mathbf{x},T)\label{eq:compacttime}
    \end{equation}
    In other words, the time dimension is compactified into a circle
    $S_1$ and the time-varying field acts as a map from
    $S_2\times S_1$ to $S_2$. This map is homotopically equivalent
    to the map $S_3\rightarrow S_2$ and since $H$ is the homotopic
    invariant, it will take on integer values which indicate the
    number of times the spacetime wraps around the two-dimensional
    sphere of the $\boldsymbol{n}$ vector. Now that the skyrmion is
    rotated by $2\pi$, the homotopic invariant is 1, i.e.
    \begin{equation}
        e^{i2\pi J}=e^{i\Theta H}=e^{i\Theta}\label{eq:spinvalue}
    \end{equation}
    that is, the spin of the skyrmion
    \begin{equation}
        J=\frac{\Theta}{2\pi}\label{eq:spintheta}
    \end{equation}

    Now the issue is quite clear. The parameter $\Theta$ controls
    the spin of the topological configuration of the
    $\boldsymbol{n}$ field, which can take on any value. Therefore
    the skyrmion can continuously interpolate between fermions with
    spin half odd and bosons with spin integer. Also possible are
    skyrmions with exotic spins like $2/3,3/5$, etc. These
    configurations will cause fractional statistics, which are often
    coined as anyons\cite{Wilczek:1982wy}.

    These discussions of the nonlinear $\sigma$-model reveal the
    topological significance of this model. Skyrmions as
    configurations with nontrivial topological charge have finite
    energy and carry fermionic number with them. Regarded as
    extended particles, they exhibit exotic spin and statistical
    properties because of the spacetime dimension. It is a quite
    rich subject and can be applied in lower-dimensional
    electromagnetic systems as well.

\bibliography{skyrmion}

\end{document}